\documentclass[a4paper,10pt]{article}
\usepackage{amsmath}
\usepackage{amssymb}
\usepackage{graphics}
\usepackage{rotating}
\usepackage{cite}
\usepackage{color}

\title{A note about the factorization of the angular part of the Laplacian and its application to the time-independent Schr\"odinger equation}
\author{Daniel Alay\'{o}n-Solarz \underline{(danieldaniel@gmail.com)}}

\begin{document}

\maketitle

\begin{abstract}
Removing al least one point from the unit sphere in $ R^{3}$ allows to factorize the angular part of the laplacian with a Cauchy-Riemann type operator. Solutions to this operator define  a complex algebra of potential functions. A family of these solutions is shown to be normalizable on the sphere so it is possible to construct associate solutions for every radial solution to the time-independant Schr\"odinger equation with a radial potential, such that this family  of solutions is square integrable in $R^{3}$. While this family of associated solutions are singular on at least one half-plane, they are square-integrable in almost all of $R^{3}$.
\end{abstract}
\section{Angular holomorphic functions}

Consider the unit sphere $S^{2}$ and its standard parametrization in spherical coordinates:
\begin{equation*}
(\cos \theta \sin \phi, \sin \theta \sin \phi, \cos \phi).
\end{equation*}
Now consider complex functions defined on almost all the sphere of the form
\begin{equation*}
u(\theta, \phi)+iv(\theta, \phi).
\end{equation*}
We impose a Cauchy-Riemann type system of equations defined by
\begin{equation}
\frac{\partial u}{\partial \theta}= \sin \phi \frac{\partial v}{\partial \phi}
\end{equation}
and
\begin{equation}
\frac{\partial v}{\partial \theta}= -\sin \phi \frac{\partial u}{\partial \phi}
\end{equation}
to which we associate the following operator
\begin{equation}
D:=\frac{\partial}{\partial \theta} + i \sin \phi \frac{\partial }{\partial \phi}
\end{equation}
let us call angular complex functions that satisfy (1) and (2) \textit{angular holomorph}. It is immediate that a function $f=u + iv$ is angular holomorph  if and only if
\begin{equation*}
D f = 0
\end{equation*}
Our motivation for studying the operator $D$ is that it allows to factorize the angular part of the Laplacian:
\begin{equation}
 \frac{1}{\sin^2 \phi }\frac{\partial^2 }{\partial \theta^2} +\frac{\partial^2 }{\partial \phi^2} +\cot \phi \frac{\partial }{\partial \phi} = \frac{1}{\sin^2 \phi}(\frac{\partial}{\partial \theta} - i \sin \phi \frac{\partial}{\partial \phi})(\frac{\partial}{\partial \theta} + i \sin \phi\frac{\partial}{\partial \phi}) 
\end{equation}

\newtheorem{prop1}{Proposition}
\begin{prop1}
Let $f$ and $g$ be angular holomorphic functions, then $fg$ and the algebraic inverse $f^{-1}$ are angular holomorph. 
\end{prop1}
\begin{prop1}
Let $g$ be a complex holomorphic function, then $g \circ f$, wherever this composition is possible, is angular holomorpic.
\end{prop1}
\begin{prop1}
Let $u + iv$ be an angular holomorph function, then $\Delta u = \Delta v = 0$
\end{prop1}
\textbf{Proof} The first proposition's  proof is straight-forward. Let $f=u+iv$ and $g=u'+iv'$, then
\begin{equation}
fg = \tilde{u}+\tilde{v}i = uu' -vv' + (uv'+u'v)i
\end{equation}
and
\begin{equation}
\frac{\partial \tilde{u}}{\partial \theta}=\frac{\partial u}{\partial \theta}u'+\frac{\partial u'}{\partial \theta}u - \frac{\partial v}{\partial \theta}v' - \frac{\partial v'}{\partial \theta}v
\end{equation}
\begin{equation}
=\sin \phi [\frac{\partial v}{\partial \phi}u'+ \frac{\partial v'}{\partial \phi}u+\ \frac{\partial u'}{\partial \phi}v + \frac{\partial u}{\partial \phi}v']
\end{equation}
\begin{equation}
=\sin \phi \frac{\partial \tilde{v}}{\partial \phi}.
\end{equation}
we also have
\begin{equation}
\frac{\partial \tilde{v}}{\partial \theta} = \frac{\partial u}{\partial \theta}v'+\frac{\partial u'}{\partial \theta}v + \frac{\partial v}{\partial \theta}u' + \frac{\partial v'}{\partial \theta}u
\end{equation}
\begin{equation}
= \sin \phi [\frac{\partial v}{\partial \phi}u'+ \frac{\partial v'}{\partial \phi}u-\ \frac{\partial u'}{\partial \phi}v - \frac{\partial u}{\partial \phi}v']
 \end{equation}
 \begin{equation}
=-\sin \phi \frac{\partial \tilde{u}}{\partial \phi}.
\end{equation}
In the same way it is possible to prove that the algebraic inverse $f^{-1}$ is angular holomorph. As a consequence we deduce then the \textbf{Proposition 2}. The \textbf{Proposition 3} is proved with a simple inspection of the laplacian in spherical coordinates and the fact that $D$ is a factor of the angular part of the Laplacian. \\
On the geometrical side it is possible to prove that if $u +i v$ is angular holomorph then
\begin{equation}
\nabla u \cdot \nabla v = 0
\end{equation}
\section{Example: The time-independent Schr\"odinger equation}
Let\ $\nu(r)$ be a real function depending on $r$, the distance to the origin in spherical coordinates. We consider the equation:
\begin{equation}
(-\Delta + \nu )f=0
\end{equation}
Where $f$ is a complex function. Suppose first that $g(r)$ a real function depending only on $r$ and that satisfies:
\begin{equation}
(-\Delta + \nu )g=0
\end{equation}
Then for every angular holomorph function, say $h$ we have:
\begin{equation}
(-\Delta + \nu )(gh)=0
\end{equation}
We illustrate this with an example. Let 
\begin{equation}
\nu := (n^2 - \frac{2n}{r}).
\end{equation}
Observe that
\begin{equation}
e^{-nr}
\end{equation}
is a solution for this $\nu$. \\
Now we construct a family of angular holomorph functions in the following manner, first we observe that
the function
\begin{equation}
\theta + i \ln \tan \frac{\phi}{2}
\end{equation}
is angular holomorph, multiplying this function by $-i$ and then applying the exponential and one obtains
\begin{equation}
\tan \frac{\phi}{2} e^{-i \theta}
\end{equation}
Now let $k,m$ be a pair of integers such that 
\begin{equation*}
 0 \leq \, \mid k \mid \, \leq \mid m-1 \mid
 \end{equation*}
 and consider the family:
 \begin{equation*}
 h_{\frac{k}{m}}:=  u_{\frac{k}{m}}+i v_{\frac{k}{m}}
\end{equation*}
defined as
\begin{equation*}
 u_{\frac{k}{m}} = (\tan \frac{\phi}{2})^{\frac{k}{m}} \cos \frac{ k \theta}{m} 
\end{equation*}
\begin{equation*}
v_{\frac{k}{m}} = - (\tan \frac{\phi}{2})^{\frac{k}{m}} \sin \frac{k \theta}{m}
\end{equation*}

Observe that
\begin{equation*}
\int_{0}^{\pi} (\tan \frac{\phi}{2})^{\frac{2k}{m}} \  \sin \phi \ d \phi = \frac{2 k \pi}{m \ \sin (\frac{k \pi}{m})}
\end{equation*}
Now we argue that since this family is obtained from a complex holomorphic function acting on a angular holomorph function then $u$ and $v$ are harmonic functions. \\
The product of one of the elements of this family and the solution depending on $r$ can be written as:
\begin{equation*}
g_{\frac{k}{m}}:= \frac{1}{\pi}\sqrt{\frac{n^3 m \sin(\frac{ k \pi}{m} )}{k} } (\tan \frac{\phi}{2})^{\frac{k}{m}} e^{-(nr +i \frac{k \theta }{m})}
\end{equation*}
Observe that
\begin{equation*}
(-\Delta + \nu)g_{\frac{k}{m}}= 0
\end{equation*}
moreover, because of the restriction imposed on $k$ and $m$ we have
\begin{equation*}
\int_{R^3} \mid g_{\frac{k}{m}} \mid^{2} dV = \int_{0}^{\infty} \int_{0}^{2 \pi} \int_{0}^{\pi} \mid g_{\frac{k}{m} }\mid^{2}  \sin \phi \  d \phi \ d  \theta \ r^{2} \ dr = 1
\end{equation*}
in almost all of $R^{3}$






\end{document}